# Traffic Road Congestion System using by the internet of vehicles (IoV)

**MUHAMMAD SHOAIB FAROOQ[1], SAWERA KANWAL[1]**
[1]Department of Artificial Intelligence, School of System and Technology, University of Management and Technology, Lahore, 54000, Pakistan
Corresponding author: Muhammad Shoaib Farooq (shoaib.farooq@umt.edu.pk)

**ABSTRACT** Traffic problems have increased in modern life due to a huge number of vehicles, big cities, and ignoring the traffic rules. Vehicular ad hoc network (VANET) has improved the traffic system in previous some and plays a vital role in the best traffic control system in big cities. But due to some limitations, it is not enough to control some problems in specific conditions. Now a day invention of new technologies of the Internet of Things (IoT) is used for collaboratively and efficiently performing tasks. This technology was also introduced in the transportation system which makes it an intelligent transportation system (ITS), this is called the Internet of vehicles (IOV). We will elaborate on traffic problems in the traditional system and elaborate on the benefits, enhancements, and reasons to better IOV by Systematic Literature Review (SLR). This technique will be implemented by targeting needed papers through many search phrases. A systematic literature review is used for 121 articles between 2014 and 2023. The IoV technologies and tools are required to create the IoV and resolve some traffic rules through SUMO (simulation of urban mobility) which is used for the design and simulation the road traffic. We have tried to contribute to the best model of the traffic control system. This paper will analysis two vehicular congestion control models in term of select the optimized and efficient model and elaborate on the reasons for efficiency by searching the solution SLR based questions. Due to some efficient features, we have suggested the IOV based on vehicular clouds. These efficient features make this model the best and most effective than the traditional model which is a great reason to enhance the network system.

## I. INTRODUCTION

The use of VANET technology and different network architectures contribute to traffic congestion in the transportation system [1]. In traditional VANET architectures, all nodes are equally important and can act as relay nodes, regardless of their location or role [2]. The routing protocol disseminates network topology information among immediate neighbors and gradually throughout the network, allowing routers to gain knowledge of the network [3]. To establish connectivity in on-demand routing protocols, routes to a node are discovered by flooding request messages[4]. Mobile ad hoc network development relies significantly on simulation as an essential tool. Simulation facilitates the study of network behaviors and characteristics under various conditions and parameters [5]. However, Simulation is not a foolproof approach to ensure the protocol's practical viability, as simulators have assumptions and simplified models that may not precisely represent real network operations.

The purpose of Ad Hoc routing protocol is to address the issue of complex network structure changes due to fast node movement [6]. In other paper proposed intelligent system for monitoring and managing roads stands out from existing systems because it incorporates three related components, namely infrastructure, transport, and cloud that interact with each other in a structurally integrated manner [7].

The research analyzed the behavioral and performance aspects of three routing protocols - proactive, reactive, and hybrid - in an urban environment [8]. This study was conducted using NS2 simulator and included the application of TCP and UDP transport protocols with different car densities. According to [9], the routing table is the primary data structure that stores all the necessary information about routes, including destination address, sequence number, hop count, next hop, lifetime, precursor list, and route state. In emergency situations and harsh weather conditions, SAR data is useful for detecting changes in the environment, as highlighted in [10]. In [11], the authors investigate the features of ad hoc routing protocols - OLSR, AODV, and ZRP and analyze performance metrics such as packet delivery ratio, end-to-end delay, throughput, and jitter under increasing node density in the network.

In [12], the authors utilize the NS2 simulator to assess the performance of AODV, DSR, OLSR, and DSDV protocols under various conditions. The Ad Hoc committee and the International Committee on Systematic Bacteriology acknowledge the Bacteriology Division of the International Union of Microbiological Societies and the International Council of Scientific Unions for their support, as mentioned in [13]. To improve the overall reliability of embedded sensor networks, [14] proposes a heterogeneous backup scheme that involves substituting one type of resource with another. According to [15], MANETs are characterized by high mobility, with new nodes continuously joining the network while existing ones move within or exit the network. Numerous ad hoc network protocol. [16] Discusses the difficulty of constructing intrusion detection systems for wireless sensor networks and mobile ad hoc networks. The paper presents an overview of current intrusion detection methods and identifies significant areas for future research. Another paper, [17] introduces the IVG (Inter-Vehicle Geocast) protocol, which employs ad hoc wireless networks to broadcast user alarm messages to all vehicles on a highway. The protocol aims to alert drivers of any potential hazards or obstacles, while also providing information on risk areas, driving direction, and vehicle positioning.

The paper introduces a system for real-time traffic monitoring and vehicle tracking in public or private transportation sectors [18]. It utilizes a social network service to provide individual users with traffic monitoring capabilities. A prototype model is developed and presented to showcase the system's functionality and performance evaluation.

In 2020, the focus of network architecture design was to achieve direct communication between vehicle-to-vehicle (V2V) and vehicle-to-infrastructure (V2I) using VANET technology and IoV [1]. However, the adoption of VANET has been limited due to insecure network service quality and device inconsistency. To address this, a smart logistics vehicle management scheme based on IoV was proposed, where smart vehicles are connected to multiple networks and share information with road-side units and other road users [19]. The uncontrolled flow of traffic in cities creates problems such as air pollution and accidents, which can negatively impact travel time and delay fraction [20]. Therefore, the accurate prediction of travel time is crucial in such situations.

Real-time traffic updates can be obtained using cameras, sensors, and AI-based predictive algorithms in traffic signal management systems [21]. Smart Transportation Systems (STS) use IoT technology and Big Data analytics to provide real-time traffic updates [22]. A sustainable and robust traffic management system can be achieved using the Internet of Vehicles (IoV) technology, which is a structured system controlled by a central device [20]. The Intelligent Traffic Light Controller Using the Rooted System is an example of an IoV-based traffic control system [23]. The use of IoT technology in traffic control systems has improved traffic management from traditional to advanced systems [24]. Contribute the IOV system through cloud-based computing to discuss better. The central control system of IOV may require expensive communication devices but through cloud computing the technique such a way that there is a leader vehicle will be selected which has all the computing and sharing resources for vehicular cloud-based computing, and another vehicle that has to store, sense traffic, measure speed, onboard unit, and other resources will be a participant in the network [25]. Discussed a comparative study to determine important reasons for Vehicular-based computing with the help of SLR. We will discuss the architecture of the modern model of IOV based on cloud computing and all challenges that may face by developers. We shall also point out some challenges that happened during the architecture of IOV cloud-based computing. These challenges are majority related to security-related issues [24].

The main structures of this paper are to all issues in the traditional traffic control system to the point of SLR. The related paper will be selected in the first phase then skip irate late pear in terms of the target paper for searching for the required answers.

The proposed system aims to address traffic-related problems using modern technology, such as IoV, MAC, DSRC, and SUMO. While VANET was a previous network model that aimed to solve traffic-related problems, it faced deficiencies in terms of security and coverage area [19]. A systematic literature review (SLR) was conducted to identify the best network for controlling traffic and to discuss the outcomes of using IoV. Machine learning algorithms have also found use in various domains, including intelligent transportation systems [26]. The proposed system will be automated and will use real-time data exchange between vehicles to visualize the city traffic on an OBU for drivers [1]. By using cloud-based competitors to control traffic systems, this system aims to deal with the issue of heavy traffic in big cities. To our knowledge, there is no such study related to IoV automated system will be beneficial for solving congestion issues.

The objective of this proposed work has to present a systematic literature review (SLR) domain in IOV and control the traditional traffic system. VANET and discussed some outcomes in terms of the best network for controlling the traffic. We have needed to solve traffic problems using modern technology IoV and two protocols MAC (media access control) and DSRC (dedicated short-range communication) and a SUMO (simulation of urban mobility) tool which is used for animating the traffic on OBU (On-Board Unit) in front of drivers for visualization of the city traffic. Which mainly uses real-time information exchange data between vehicles. This modern system is automated which will be beneficial for solving traffic-related problems using SLR-based questions.

The structure of this paper is outlined in [18]. Section 1 provides a brief introduction to the development of IoV and highlights the significance of collecting, summarizing, analyzing, and categorizing current research in this area. In Section 2, the background of IoV is presented, including a systematic literature review and a comparison to previous work. Section 3 describes the research methodology, including research questions, presence/prohibiting criteria, and search strings used to gather relevant studies in the IoV domain. The research results are presented in Section 4 through tables, graphs, and a proposed IoV model. Section 5 offers a summary of the findings via a research hierarchy and a designed IoV-based smart model. In Section 6, open issues and challenges in IoV are discussed from various perspectives, based on selected papers. Finally, research legitimacy threats are presented in Section 7, while Section 8 concludes the article.

## II. RELATED WORK

The review found that the majority of the studies on the use of IoV for traffic road congestion management have been focused on the use of vehicle-to-vehicle (V2V) and vehicle-to-infrastructure (V2I) communication. Crowding can also be described as an increase in the expenses of road users resulting from the disturbance of regular traffic flow [27]. Some restrictions in term of network security, reliability and consistency. The integration of IoT in the IOV-based traffic control system is essential for the efficient management of city traffic [28]. However, security and privacy measures in IoV pose challenges that need to be addressed [29]. GPS and other location-based technologies can provide drivers with useful information, but VANET faces challenges similar to those of MANET, and its wireless nature makes it less secure than wired networks [30]. The dynamic nature of VANET poses challenges in designing routing protocols that can support fast-moving vehicle nodes [31, 32]. AODV-MEC, a novel approach for clustering AODV routing, utilizes edge computing techniques to overcome these challenges [33]. VCC-based traffic management systems have been reviewed and compared to VANET-based systems in terms of their effectiveness in managing road traffic [34]. As more vehicles become connected and transfer data through IOV and other infrastructure technologies, the need for an efficient traffic control system continues to grow [34]. Various standards and protocols have been developed for vehicular ad hoc networks (VANETs) and the Internet of Vehicles (IoV) to enable intelligent transport systems and remote vehicle control [24, 30, 35]. However, privacy and security concerns and data storage management need to be addressed in these networks [36]. V2V and V2I communication are used in VANETs, where vehicles communicate with other vehicles and Road Units (RSUs) [34]. A proposed framework called "vehicular cloud" aims to reduce the workload and power consumption of centralized cloud systems by utilizing computational resources from nearby vehicular clouds [37]. This framework operates in proximity to traffic lights and employs a mathematical model called "MILP" to analyze distributed task assignment compared to a single task assignment approach. A systematic review of blockchain applications in ITS and IoV can provide an overview of current research in this area [20]. The study proposes a framework that involves creating an obfuscation zone on a map, identifying the locations of users, and excluding location-based services that pertain to lakes [38].

This study introduces a new platform for vehicular data storage and processing using cloud computing and IoT technologies, which includes intelligent parking and vehicular data mining cloud services [39]. Other research focuses on developing affordable sensor systems for detecting and classifying vehicles [40], and real-time traffic control based on distributed computing platforms [41]. The Internet of Things (IoT) can be utilized for traffic management by connecting all traffic components through sensor devices [42]. Various algorithms and protocols have been proposed for improving traffic flow, such as cooperative adaptive cruise control, platooning, and traffic signal control, and vehicular ad hoc network (VANET) using DSRC communication technology has been considered for use in connected and autonomous vehicles [43]. Additionally, several studies have examined the impact of IoV on traffic congestion [27, 35, 43, 44]. These studies have shown that the use of IoV can lead to significant improvements in traffic flow and travel time. However, there are also concerns about the privacy and security of the data collected by the IoV system.

Cloud computing is one more trending, modern and compatible technology in almost every field of life. Cloud computing is basically used for use resources which may expensive for individuals like best computing, sensing and storage capacity to build large and efficient system [25]. The use of Cloud computing in IOV is that every vehicle cannot have such powerful and needed requirements to compute complex algorithms by every vehicle Overall, IoV has the potential to enhance traffic road congestion management, as well as provide safety, security, confidence, and comfort to passengers and drivers. VCC aims to offer cost-effective services to drivers while reducing accidents and traffic congestion[24]. The underutilization of communication, storage, and computing resources in vehicles is a motivating factor.

.A meaningful combination of these resources will have a major impact on society [18]. However, additional research is mandatory to address the test associated with implementing IoV, such as privacy and security.

IoT-based traffic management solution using sensors like IR and RFID, to guide ambulance drivers and identify traffic violations [45]. A framework to transform normal cities into smart cities by leveraging ICT and automating processes. The paper also considers a technique to measure traffic density for smooth vehicle movement.

Main challenge addressed in this research is the adaptation of the Public Key Infrastructure (PKI) architecture, traditionally used in wired networks, for use in Vehicular Cloud Networks (VCN). And to provide a security solution for VCNs [46]. The study comprises three steps: firstly, conducting a network architecture study to identify the key components of the network; secondly, proposing a security solution architecture; and lastly, programming and validating the solution through simulation testing.

Smart traffic management systems utilizing queue detectors installed beneath the road surface can detect traffic congestion and transmit data to a central control unit [21].

Real-time traffic monitoring and vehicle tracking systems have been designed for public and private transportation sectors, leveraging social network services to provide individual users with traffic monitoring [18]. Various strategies and automated sensor systems have been proposed to analyze traffic density and alleviate congestion, with a review of different sensor systems evaluating their advantages and drawbacks [47]. An intelligent traffic monitoring system is proposed utilizing global unique EPC codes and RFID readers for all-weather operations [48]. VANET faces security issues, while IoV, which is dependent on IoT and cloud computing, offers a self-organized, infrastructure-less, and automated system for traffic control that can service larger areas [20, 49, 50]. The use of Internet of Vehicles (IoV) has been proposed as a solution to traffic congestion management, and a majority of research has focused on utilizing vehicle-to-vehicle (V2V) and vehicle-to-infrastructure (V2I) communication to improve traffic flow. However, IoV-based vehicular communication faces security threats, categorized as Inter-vehicular attacks and intra-vehicle attacks. Inter-vehicular attacks involve malicious nodes that can attack communication to divert actual information, resulting in potential life and property loss. This type of attack is known as masquerading or impersonation attack. On the other hand, intra-vehicle attacks can target all participating components of a car by any criminal. If the sensors within a car are attacked for criminal purposes, it may result in compromising driver safety and vehicle integrity. [51] Meanwhile, significant technological advancements in the transportation sector have led to the implementation of computer vision in Intelligent Transportation Systems (ITS) to analyze video data sourced from CCTV cameras [52].

In other discuss shows that big data can play a key role in provided that sound and valuable predictions and also provide a comprehensive analysis of several methods, tools, and techniques for the use of big data in IoV [53]. The literature also suggests that the use of IoV can lead to significant improvements in traffic flow and travel time but there are also concerns about the privacy and security of the data collected by the IoV system.

In the table 1 analysis all those points which is used in this research as referenced and after that deals all problems in traditional vehicular ad hoc networks.

Our proposed study aims to compare and analyze the differences between the IOV based on cloud network and traditional VANET models, highlighting their underlying architectures and communication mechanisms. The IOV model relies on cloud computing and communication technologies to enable real-time data exchange between vehicles, infrastructure, and the cloud, while VANET uses direct vehicle-to-vehicle communication. The study will explore the advantages and limitations of both models and address any existing problems to develop an optimized and efficient network model.

**Table 1**: Comparatively analysis between literature and purposed work.

| Sr. | Article Name | Important points by Literature | Features of Purposed system |
|---|---|---|---|
| 1 | VANET architecture and protocol stacks: A Survey | An Infrastructure-less system, low covered area | Infrastructure based and covered more area like city range. |
| | | No real time Information sharing | Share Real time information to network by automatic and intelligent system |
| 2 | Systematic literature review on Internet of vehicles communication security | Deal about Inter-vehicle attacks and intra-vehicles attacks | Discuss about the possible system which minimize threats |
| 3 | Baofeng; Li, Chungue Survey on the Internet of Vehicles: Network Architectures and Applications *IEE*, 2020. | Basic building block of IoV and comparison between VANET and IOV | Enhancement the concept of IOV using vehicular clouds and integrate with modern needs |
| 4 | Network architecture in Internet of vehicles (IoV): Review protocols, Analysis, challenges and issues. | Discussion about routing protocols and their challenges used in IOV | Elaborate the threats in existing routing protocols and try to overcome it |
| 5 | Vehicular Cloud Networking:Architecture and Design Principles. IEEE Communication magazine | Discuss the Traffic control system to switch from traditional to Vehicular cloud computing | Enhancement the Vehicular control system by integrating the 5G technology to it for faster communication |

## III. RESEARCH METHODOLOGY

A comprehensive procedure for collecting and resolving main a[54]rticles is provided by the SLR for all readings in the field under consideration. In this paper, we describe a design for an IOV system that is based on cloud computing-based VCC vehicular cloud computing. This is a suitable and meditative review process that involves the following steps: 1) Define the work and objectives 2) Discuss the research questions 3) plan the exploration of research papers 4) analyze the articles for information 5) sort the articles according to keywords 6) extracting the information from the articles.

### A. Research Objectives

The major aims and objectives of the purposed work are as follows.

- ➢ To assess the potential for modern vehicular technology to address traffic problems and to determine its effectiveness as a means of traffic control.
- ➢ Identify the types of problems commonly faced in traffic control without the use of modern vehicular technology.
- ➢ Evaluate the impact of cloud computing on the Internet of Vehicles in controlling the traffic system.
- ➢ To investigate the impact of traffic congestion on driving performance and determine the extent to which it affects the driver.
- ➢ To identify the technological advancements that make the Internet of Vehicles superior to the traditional traffic system, and to assess how these improvements enhance the efficiency and effectiveness of the traffic system.

### B. Research Questions

Our field can only operate efficiently if we set our goals as research questions.
Further, a complete search design obligatory in the review for the documentation and abstraction of the greatest important articles has been conventional. Table 1 discussed the Questions to motivate that makes a speech and replies in the light of a well-defined process assumed in [30, 54]. We have set all research questions which fulfilled our purposed work in the table 2. After that we will try to collect data and analyzed them to purposed goals. The Table 2 mentioned all Research questions as research aims and objectives. It point out traditional network their positive outcomes, issues in this network, technological improvements required for effective model and try to discussed architecture procedure for development the enhanced network.

**Table 2**: Research Questions and Major Motivations

| Symbol | RESEARCH QUESTION | Major Motivation |
|---|---|---|
| RQ1 | What type of problems were faced during the usage of traditional Vehicular ad hoc networks in literature? | To explore all problems faced using traditional traffic control systems by the drivers discussed in the literature |
| RQ2 | What type of traffic issues in big cities can be solved with IoV in literature? | To get knowledge about those issues which can be solved by the IoV Architecture model discussed in the literature. |
| RQ3 | Which technological improvement makes IoV better than VANET? | Identify all those technological enhancements which make our IoV architecture better than the old traffic control system. |
| RQ4 | What are the research types to explore IoV for solving Traffic problems? | To Explore all those journals which published most research articles to enhance the traffic control system. |
| RQ5 | Which steps should be taken to deal with the challenges of traffic congestion in the Vast Area? | To explain the procedure of Vehicular cloud computing-based IOV architecture and deal with its challenges. |

## C. SEARCH STRATEGY

Our last step in SLR is to research the best, most relevant, and latest articles about our latest work. In this process, the use of the search string technique is very important because needed paper search by easy by the efficient search string. The effective search string cannot search the next phases and will also be inefficient. Many features of composing articles related to fields compose many perceptions related to the study.

## D. SEARCH STRING

Discussed all search strings which we have used for collect the targeted articles. Search string selection is very important to search the most suitable and targeted articles for research work in Table 3. By using numerous search strings and a keyword built on the internet of vehicles, we have demonstrated that the keyword was in force and reasonable. We use the search string to find relevant papers about our research objectives. We should also use alternative keywords (Synonyms) to describe a wide range of research areas. Alternative keywords can be found in abstracts, indexes, and on the internet. By using the final selected keyword and its alternative keywords, we will find the most relevant, latest, and validated articles. Through this process, a search string can be formulated for related articles by putting them in relevant online directories.

These are technologies used in vehicular cloud networks Communication and transmission. The finalized search string has three fragments. As we have discussed earlier that VANET is also used now a day to overcome traffic congestion problems. This system also gets positive results to resolve traffic congestion problems. VANET model operated by (V2I) vehicle to infrastructure, (V2V) vehicle to Vehicle, and V2P vehicle to pedestrians. This system has worked like MANET mobile ad hoc network which is infrastructure-less and without a central controlling device. VANET is a branch of MANET which follow the rules of MANET. But this system may have some problems happened in the model. Like this problem is not automated because all nodes and vehicles in the VANET model communicate with each other only in the real-time system and only shares information with a neighbor in the vehicle's roadside unit and On-Board Unit. This system is not able to share all information related to traffic congestion in the whole city. Due to these reasons, all problems cannot resolve efficiently.

## E. LITERATURE RESOURCES / JOURNAL REPOSITORIES

The table 4 mentioned the publishers that have published the articles selected for the systematic literature review (SLR). It has also mentioned the prominent articles name corresponding to Repository. The exploration was approved to collect data using the subsequent search terms: (''IoV'' OR ''VANET'' OR ''Connected Cars'') AND (''IoT OR ''Vulnerability ''OR ''ITS).

**Table 3:** Terms and keywords used in the search

| TERMS (KEYWORDS) | Synonyms/ Alternative strings |
|---|---|
| Internet Based (IB) | Internet of Vehicles (IoV) and Internet of Things (IoT). ITS |
| SUMO | Simulation of urban mobility animation tool (SUMO), |
| VANET, MANET | Vehicular ad-hoc network and Mobile ad-hoc network. |
| DSRC | (Dedicated short-range communication). MAC(media access control), |

Table 4: Publisher-wise Search strings

| Repository | Search Strings |
|---|---|
| Science Direct | (Social internet of vehicles OR Social internet of things OR Intelligent Transportation system AND Targeted, Enabling IOV technologies) |
| SCIENCE DIRECT | (Systematic Literature Review OR SLR AND Internet of Vehicle Communication security OR IOV authenticity OR IOV Integrity OR IOV Confidentiality). |
| IEEE Xplore | (Blockchain technology OR Automotive communication AND Intelligent Transportation system OR Internet of vehicles AND SLR OR systematic literature review). |
| IEEE Xplore | (Smart logistics management OR Smart logistic transportation AND Internet of vehicles OR IOV OR connected vehicles.) |
| IEEE Xplore | (Computation-intensive graph jobs OR vehicular computationally intensive graph AND vehicular clouds AND IOV OR internet of vehicles.) |
| ACM | (Survey AND Road traffic congestion OR vehicular congestion AND traffic measures OR Transportation analysis OR road evaluation AND sustainable and Resilient transportation.) |
| ACM | (Vehicular cloud networking OR VCN OR Transportation cloud networking OR intelligent vehicular cloud computing AND architecture OR Network Design AND principles OR standards.) |

## F. INCLUSION AND EXCLUSION CRITERIA

Constrictions distinct for addition criteria (AC) are:
  **IC1)** Consist of readings that remained used mainly showed for the internet of vehicles cloud-based network.
  **IC2)** if the study was directing the IoV cloud base network enhancing traffic system.

The elimination criteria were functional to all the articles except readings that involved a Vehicular ad-hoc network

(VANET) that affects or elaborates on other extraneous processes,

    **EC1)** if the study did not contain any expectations and only include IoV enhancement cloud-based network.
    **EC2)** doesn't have any discussion VANET computational study.
    **EC3)** if the study involved IoV cloud-based vehicle system.

### G. SELECTION OF RELEVANT PAPERS

The first phase is the studies based on labels and replication elimination. This article was obtainable but remains inappropriate in the selected domain, in selection persistence, abstracts were cautiously studied articles were involved that describe the computational certain area. Furthermore, next to investigative the articles follow inclusion criteria. Lastly, the selected IoV domain articles remained included in the following valuation level.

### H. ABSTRACT-BASED KEYWORDING

In this phase, the screening process is implemented on Abstract base keywords. We have to analyze the abstract base reading and screen the relevant keywords from the abstract. After analyzing them we can decide on the current paper to target our objective work. The primary idea can easily get by analyzing the abstract base reading so it is very beneficial in targeting our papers. If keywords are found in this phase, the next work is to combine these keywords to search for relevant work.

The designated articles have been divided into four major types; IEEE, Science Direct, and ACM. Other domains like the internet of things (IoT) and the social internet of vehicles (SIoV). Furthermore, just for clarity IEEE explore has several recognized IoV and VANET [1], [6,7] considered and grouped here as Science Direct.

### I. ASSESSMENT CRITERIA FOR QUALITY

The evaluation of the systematic quality of the targeted articles is an essential task of a systematic review. We know that all studies of given articles are different from each other according to their design, purpose, and technique of research but the process of evaluation should be according to the sequential assessment process suggested [30]. The whole process should be according to the assessment standard set for the evaluation of the quality. Every step has its standards to evaluate and plan to rehear the standards of objectives, methodologies, and ways to collect the analyze them, and their results should be followed [51, 55]. The survey will be administered using articles written by various authors, as described below. Subsequently, distinct approaches will be utilized to answer each questionnaire. The internal and external quality standards may also be employed to assess the caliber of systematic reviews [55].

 We have been conducting an online survey for the population who can respond to our questionnaire and after that we have been analyzing them by explain the survey responses according to the literature review for the evaluation of systematic quality in given table 5.

Table 5: Questionnaire to assess the quality

| Sr. | The assessment Questions | Expected Answers | Score |
|---|---|---|---|
| | | **Internal Scoring** | |
| 1 | Can we control the traffic problems with modern vehicular technology? | Yes | 86.2% |
| | | No | 13.8% |
| 2 | Which type of problems are mostly faced without modern vehicular technology? | heavy congestion | 32.1% |
| | | Road congestion | 53.6% |
| | | wastage of time | 14.3% |
| 3 | What is the effect of cloud computing based on the internet of vehicles for controlling the traffic system? | better usage of resources | 53.6% |
| | | expensive cost | 25% |
| | | less cost | 1% |
| | | not know | 17.9% |
| 4 | Does traffic congestion affect your performance during driving? | Yes | . 67.9% |
| | | No | . 17.9% |
| | | May be | 14.3% |
| 5 | Which technological improvements make vehicles better than traditional the traffic system of our country? | traffic resources by cloud system | 28.6% |
| | | intelligent traffic control system | 64.3% |
| | | cover large area | 7.1% |

## III. DATA ANALYSIS

In this chapter, we have concluded the technologies IOV (internet of vehicles) and configuration for tools SUMO (simulation of urban mobility). These technologies aid in managing vehicular traffic and optimizing urban mobility. IOV refers to the integration of vehicles with the internet, allowing for real-time communication between vehicles and infrastructure, while SUMO enables simulation-based testing and optimization of transportation systems. A detailed study is needed to answer the question of SLR. The Table 6 explore selection process of targeted SLR related to our Goal. It consists of four phases where first phase selects the articles after the input the selected strings in articles repository. After that the unrelated articles has removed on the basis of analysis most suitable SLR. First of analyzed the review answer by population and then Analyzed the SLR questions conducted to achieve the objective of our work.

### A. SEARCH RESULTS

There are 121 articles found during the primary stage of the article search. An initial search has been explained and it involves searching all articles using specific keywords chosen for the search. These articles have been searched from online repositories mentioned in the Table#6. After these four phases are applied to the target the relevant articles are issued for extracting the data.

**TABLE 6. Publisher Based Stage Wise Selection Process**

| Database/Repository | Primary Search | P-I | P-II | P-III | P-IV |
|---|---|---|---|---|---|
| IEEE EXPLORE | 50 | 26 | 11 | 10 | 5 |
| SCIENCE DIRECT | 25 | 12 | 8 | 5 | 3 |
| ACM | 26 | 8 | 7 | 5 | 4 |
| SPRINGER LINK | 50 | 20 | 16 | 8 | 5 |
| **TOTAL** | **121** | | | | **17** |

First of all title based selection is done after the primary search in which the title-based selection of articles is implemented called Phase 1 of the selection process. In this phase, all those artists not included in the first case will not be selected for the next phase. The selection is done by analyzing the abstract. Data extraction is performed after a detailed study of the paper. final 17 papers are collected as Phase 4th stage in which detail of each repository is given in table number 5.

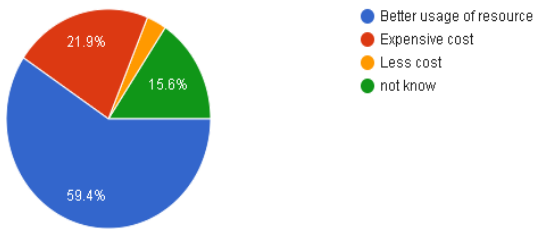

Figure 1: impact of cloud system on IoV

### B. ASSESSMENT AND DISCUSSION OF RESEARCH QUESTIONS

In this section, we will evaluate and analyze the research question, and then assess the findings based on a review of population analysis. We conducted an online survey using a Google form to collect positive responses and targeted specific audiences for valid responses. This process involved reviewing all relevant articles to analyze the research questions.

### 1) WHAT IS THE IMPACT OF CLOUD COMPUTING ON TRAFFIC CONTROL SYSTEMS, SPECIFICALLY WITH REGARD TO THE INTERNET OF VEHICLES?

It is most important to set the domain of reading articles in terms search the targeted data for review. We have selected the articles for this purpose from January 2014 to June 2023. We have selected 8 articles in terms of targeted and needed data. We have conducted online survey based on questions, the review graph is given below by the selected population and we have analyzed this section to findings reviews. The majority of reviews have been reviewed on better usage of resources 59.4% expensive cost 21.9% and not know is 15.6% in figure 1. It means that this point is valid the cloud computing-based IOV has control of the traffic due to better usage of resources.

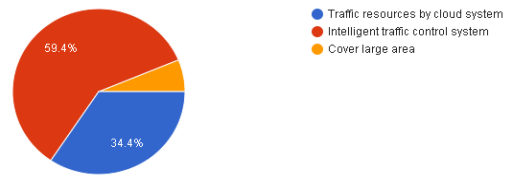

**Figure 2:** problems are faced without IoV technology

### 2) WHICH TYPE OF PROBLEMS ARE MOSTLY FACED WITHOUT MODERN VEHICULAR TECHNOLOGY?

The transportation industry may encounter significant problems in the absence of modern Internet of Vehicles (IoV) technology. Some of these issues include limited connectivity between vehicles, infrastructure, and other devices, resulting in ineffective communication and coordination. This can lead to traffic management inefficiencies, reduced safety, and longer travel times. In this question base result of majority reviews have been reviewed on intelligent traffic control system and responses of 59.4%. Traffic resources by cloud system responses of 34.4%. It means that this point is valid IoV is intelligent traffic control system in Figure 2. In addition, the absence of real-time data can make it difficult to accurately predict traffic patterns, congestion, and other crucial information. This can impede route optimization, emergency planning, and traffic flow enhancement. Furthermore, without modern IoV technology, various safety features like advanced driver assistance systems (ADAS) and collision avoidance systems may be limited or unavailable, which can increase the risk of accidents and fatalities.

Increased Environmental Impact: Without IoV technology, vehicles may not be optimized for efficient driving, leading to higher fuel consumption and increased carbon emissions.

Inefficient Fleet Management: Without real-time data on the status of vehicles, it can be difficult to manage fleets efficiently. This can result in increased maintenance costs, longer downtime, and reduced productivity.

An IEEE "Institute of Electrical and Electronics Engineers." IEEE vehicular technology society section 1st March 2022. It has been contributing a lot of concepts terms of controlling the traffic system with modern vehicular technology. It has also found gaps between the current system and modern needs. It has also contributed to the challenge is in modern vehicular system. We have many articles found by different online repositories used also uses them in this work [49].

Some paper explores this novel architecture of (IoV) cognitive internet of vehicles, as well as research occasions in a vehicular ad-hoc network. We have presented present an overview of IoV including its evolution, related technologies, and architecture [43].

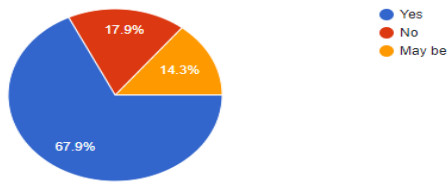

Figure 3: Traffic control system using by IoV

### 3) CAN WE CONTROL THE TRAFFIC PROBLEMS WITH MODERN VEHICULAR TECHNOLOGY?

This study aims to reduce congestion and messaging overhead, and we can control traffic problems using some techniques, tools and modern technological concepts.

Intelligent Transportation Systems (ITS) technology can help manage and control traffic by providing real-time information to drivers, such as traffic conditions, alternate routes, and estimated travel times. ITS also includes systems that can automatically adjust traffic signals and manage traffic flow based on current conditions. Connected vehicles use wireless communication to share data between vehicles and infrastructure. This technology allows vehicles to communicate with each other and with traffic management systems, which can help reduce congestion and improve safety on the roads. Autonomous Vehicles: Autonomous vehicles have the potential to significantly reduce traffic problems by eliminating human error, reducing congestion, and improving fuel efficiency. Self-driving cars can communicate with each other and traffic management systems to optimize traffic flow and reduce traffic jams.

After creating an online survey form and sharing it with all concerned stakeholders, we collected reviews from people which were predominantly positive. The majority of reviews have been reviewed on Yes 67.9%, No 17.9% and May be 14%. It means that this point is valid many traffic problems control with IoV in Figure 3. It was found that the traditional system has limitations, such as the lack of real-time traffic sharing with other nodes and sharing only among neighboring vehicles, which is insufficient to control the traffic of an entire city. In contrast, our enhanced system is designed to address these issues by incorporating Internet of Vehicles (IoV) technology and integrating long-range communication nodes, along with using Vehicular Cloud Computing. This allows our system to have better long-range traffic control capabilities compared to the traditional system.

Brings vision into the covered architecture of the social internet of vehicles (SIoV) by demonstrating the role and architecture of each entity of the system along with permitting technologies and protocols. Another main contribution of this article is to highlight the social relationships between different objects of the system along with the management dynamic nature of SIoV systems. And analyzes the proposed SIoV architecture by representing separate use cases and pronouncing the feasibility of the proposed architecture [55] The architecture of IoV Design positions numerous challenges like reorganization, scalability, security, privacy, consciousness, heterogeneity, and interoperability.

### 4) DOES TRAFFIC CONGESTION AFFECT YOUR PERFORMANCE DURING DRIVING?

Traffic congestion can affect a driver's performance while driving. It can increase stress levels, fatigue, and frustration, which can negatively impact a driver's ability to focus and react to changing driving conditions.

When a driver is stuck in traffic, they may become impatient and try to make risky maneuvers such as changing lanes frequently or aggressively accelerating, which can lead to accidents. Moreover, stop-and-go traffic can lead to more wear and tear on the vehicle and increase fuel consumption.

Therefore, it is important for drivers to remain calm and focused when driving in traffic congestion. Drivers should allow extra time for their journeys, follow traffic rules and signals, and avoid distractions such as texting or using a phone while driving. Additionally, it may be helpful to use alternative transportation methods such as carpooling or public transportation to reduce traffic congestion on the

Traffic congestion can have an impact on your driving performance. When traffic is congested, it can lead to slower driving speeds, more frequent stops and starts, and increased driver frustration, all of which can negatively affect a driver's performance. Yes, traffic congestion can definitely affect your performance during driving. When you are driving in heavy traffic, you have to constantly pay attention to the cars around you, adjust your speed, and change lanes frequently. This can be mentally and physically exhausting, which can affect your ability to focus and react quickly to changing road conditions. In addition, driving in heavy traffic can also be stressful and frustrating, which can impact your mood and mental state. This can lead to a decrease in your ability to make good decisions and respond appropriately to unexpected situations on the road. Furthermore, being stuck in traffic for long

Periods of time can lead to physical discomfort and fatigue, such as muscle tension, back pain, and eye strain. All of these factors combined can ultimately lead to a decrease in your overall driving performance. Therefore, it is important to take breaks and practice safe driving habits, such as maintaining a safe following distance, staying alert and focused, and avoiding distractions while driving, in order to reduce the negative effects of traffic congestion on your driving performance. In figure 4 clearly mentioned that many problems may faces due to absent of modern Internet of Vehicles which can handle by using modern vehicular system reviewed on online survey "Yes" responses 86.2% and No 13.8% responses.

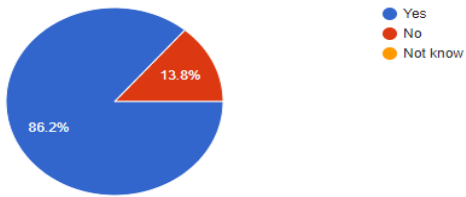

**Figure 4**: Affected your performance

## IV. DISCUSSIONS

In this section, we have going to elaborate on Review Questions. The aim of my work is about selected SLR the terms of the best solution to deal with road congestion issues in big cities. We have read about traditional traffic control systems for traffic control. But there were a lot of issues in that system. Some countries do not apply any technological system for the traffic control system.

### RQ1. WHAT TYPE OF PROBLEMS WERE FACED DURING THE USAGE OF TRADITIONAL VEHICULAR AD NETWORKS?

There most countries use traditional traffic control systems. Just a traffic signal system is not enough to control all problems related in previous decades the research has been growing the technology of traffic control. VANET (vehicular ad hoc network) is one of the most famous traffic control systems in previous decades. VANET is on the relay based on NET (Mobile ad hoc network). MANET is an infrastructures-less and self-controlled communication network in which participating nodes communicate each other without central communicating device. VANET is work same as MANET system in which all vehicles work as mobile nodes. But there are many issues also faced by drivers using VANET system all vehicles just communicate Vehicular to vehicular (V2V), Vehicular to infrastructure (V2I), and Vehicular to pedestrians (V2P). All information can get through on board unit (OBU) on front of drivers. However, VANETs has not been widely used and has not brought greater commercial value for long time [1].

#### A. NO REAL TIME TRAFFIC INFORM,ATION
This system could not able to share real time traffic information. It means that this system may not able to current situation of another place of city.

#### B. LIMITED DISTANCE
This system has a limited geographical coverage, which is one of its disadvantages.

#### C. NOT INTELLIGENCE
The system is not an intelligent one, meaning that it cannot predict potential traffic congestion based on current parameters.

#### D. SECURITY ISSUES
VANETs are vulnerable to unauthorized access, where attackers gain access to the network without permission. This can lead to the interception of sensitive data and the unauthorized control of vehicles. There may be malicious node may attack on other nodes. Other integrity, authenticity and confidentiality related issues may be found in this system. Since VANET are evolved from MANET the vulnerabilities posed by VANET are largely inherited from MANET ad hoc architecture [51].

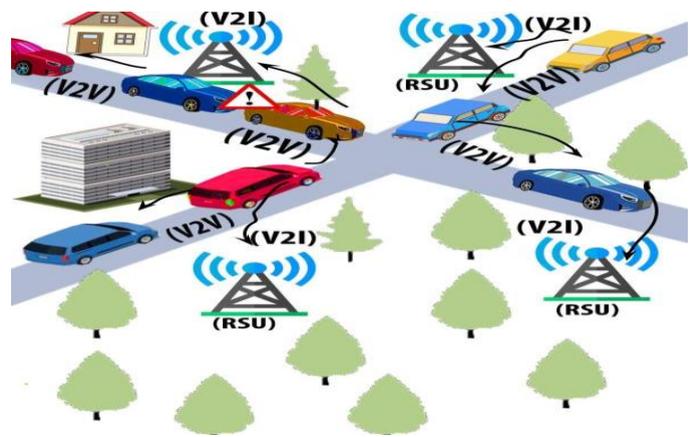

**FIGURE 8:** VANET Architecture

The figure 8 explain detailed architecture of VANET. There are many stockholders in these architectures to share information among different vehicles. The information may be vehicular to vehicular (V2V), vehicle to road side units (RSU), and vehicle to pedestrians. But this architecture have many problems like no central control system, there is no IoT Technology involve in it so real time and vast range communication cannot be achieved it is just for local level communication without intelligent system. Just one node share its information to neighboring vehicle of traffic information through MANET.

### RQ2. DESCRIBE TRAFFIC ISSUES IN BIG CITIES WHICH CAN BE SOLVED WITH THE HELP OF IOV?

There may many traffic related issues which we can solve with the help of IOV. In this section we elaborate that which traffic related issues can be solved by cloud based IOV.

- IOV solve many issues which may not able to solve with VANET. If there may a congestion in specific junction of road then our system must be able to inform all those vehicles which are coming toward conjected road. So that may follow another alternative route to reach their destination.

- This task can completely get by IOV based system.
- If anyone violate speed limit rule then this may inform the related team to control this violence.
- If road accident is occurred on specific road, then this system can inform rescue teams.
- They provide efficient time management system for traffic control. Traditional traffic signal system work on the fixed time slots to allow for cross the road. But our new system can manage this efficiently with detection and sharing the information of traffic.
- It covers large geographic area for traffic control system. It not only communicates node to node communication but also everything participating in the network like IoT.
- It has a smaller number of security and integrity Issues due to best routing protocols control by central communication system.

### RQ3. WHICH TECHNOLOGICAL IMPROVEMENT MAKES IOV BETTER THAN VANET?

In this section we elaborate with the help of related articles that which technological enhancements makes IoV better than traditional VANET system. Improvements are given below.
- IOV system is based of internet of things so it is an
- Intelligence system. In IoT all participating nodes works collaborative manner in term of perform the task. IOV is one of them application of IoT. Capability and the development of cloud computing And AI technology enable vehicles to autonomously choose to access better performing networks to insure table network connectively [1]. Specific List of Data sources directories in table 7.
- In this system all devices are compatible and collectively works according to the needed information for the traffic control.
- SUMO Simulation of urban mobility makes this
- System efficient and fast. This is a tool to process and generate traffic simulation with the help of raw form of data from sensors in the vehicles and RSU.
- IOV also have GPS to share and get information from vehicular clouds to cover vast area of traffic network.
- This system is also given facilitate to those vehicles which have not resources to compute and Storage and cloud resources. All vehicles having on bored unit and installed this system may participate and get facilitate from this network [10].

### RQ4. WHAT ARE THE RESEARCH TYPES TO EXPLORE IOV FOR SOLVE TRAFFIC PROBLEMS?

Various research types that can be used to explore the use of IOV (Internet of Vehicles) for solving traffic problems. Some of these research types include: Experimental research involves conducting experiments and collecting data on the use of IOV technologies in real-world traffic scenarios. This can help to identify the effectiveness of various IOV solutions in reducing traffic congestion and improving overall traffic flow. Survey research collecting data from individuals and organizations through surveys to understand their perspectives on the use of IOV technologies for solving traffic problems. This can help to identify the challenges and opportunities associated with implementing IOV solutions in different contexts.

Case study research conducting in-depth analysis of specific cases where IOV technologies have been used to solve traffic problems. This can help to identify best practices and lessons learned that can be applied in other contexts. Comparative research comparing the effectiveness of different IOV solutions in solving traffic problems. This can help to identify the most suitable solutions for specific contexts and provide guidance for decision-making.

Table 7: List of Data sources directories

| Sr. No. | Name of Data Source | Data Source Link | Accessible |
|---|---|---|---|
| 1 | ACM | Systematic literature review on Internet-of-Vehicles communication security | Yes |
| 2 | ELSEVIER | Social Internet of Vehicles: Architecture and Enabling Technologies | Yes |
| 3 | IEEE Explore | Blockchain Technology for Intelligent Transportation Systems: A Systematic Literature Review | Yes |
| 4 | SAM AT | Authorization Framework for Secure Cloud Assisted Connected Cars and Vehicular Internet of Things | Yes |
| 5 | SPRINGER | Network Architectures in the Internet of Vehicles (IoV): Review, Protocols Analysis, Challenges, and Issues | Yes |
| 6 | Research Gate | review article big data processing and analysis on the internet of vehicles: architecture, taxonomy, and open research challenges | Yes |
| 7 | Google Scholar | A Survey of Trust managemnet in internet of vehicles. | Yes |

### RQ5. WHAT IS THE PROCEDURE TO CREATE THE VEHICULAR BASED IOV ARCHITECTURE AND THEIR CHALLENGES?

In this section complete procedure of designing the IOV architecture based on vehicular cloud network. And also deal all challenges about this architecture.

#### A. CREATE THE CLOUD

Source allocation is initial procedure of vehicular cloud computing. The main vehicular node is selected by system which is used for manage the whole network. Head or leader node send the resource allocation message to all participants of network, this is called RREQ message.

The participants of network of network will be replied on the bases of needed information and scenario. After transfer the RREQ message to whole network the route reply message will come back head node which is responsible of mange the networks. Other nodes are just participating in the network for just get the traffic related information.

### B. GOAL ASSIGNMENT AND COLLECTION THE OUTPUT

When cloud leader receives the resource reply message (RREPs) from other vehicles then leader select of them as cloud member. The selection procedure is based on the cloud range and the correct operation of the resources that will run the application. Once the cloud members have been selected, the cloud leader assigns them to their respective tasks for collecting traffic-related data. Each cloud member then collects data from their environment and sends it to the cloud leader. If capacity of processing 50 and data storage resources of cloud leader is low then cloud leader may assign this task to other cloud member which can processed and store the task. The cloud leader also maintains the table in which all information related to vehicle and resource will be maintained. The leader also updates this table according to scenario.

### C. SHARING THE PROCESSED RESOURCES RESULTS

After processing and storage, the resources result the leader vehicle is also responsible to share this information to whole nodes of network. The connected participated nodes and concept of shared resources among all nodes. The vehicles which have needed any information about the traffic, it will send the query about it and then cloud leader send this content to this node. This is the main benefit got by vehicular cloud computing because all nodes do not have all required resources due to expensive resources. The main idea behind this concept to get all raw data is collected by cloud participated nodes which are able to collect and after that share it to leader node for computing the raw data of traffic. After this all information share among all participated vehicles in the network. Publishing research papers or articles in academic journals or other publications to disseminate the results of data processing and analysis to a wider audience. Sharing processed resource results through open data platforms, such as data.gov or Kaggle, can enable others to access and analyze the data, facilitating collaboration and innovation. Social media Sharing processed resource results through social media can increase visibility and engagement with stakeholders and the wider community. Overall, sharing the processed resource results is crucial to ensure that data-driven decisions are based on accurate, reliable, and transparent information. By using a variety of methods to communicate the results to different audiences, stakeholders can make informed decisions that have a positive impact on their organizations and communities.

### D. MAINTENANCE OF VEHICULAR CLOUDS

All vehicular nodes work in the area of cloud range and after this it will leave the network. When a cloud member wants to leave the network the cloud leader searched another node which will perform this resources task. The leader again send message to whole network for assign the task then vehicular nodes that can take this task will send request reply to cloud leader. Then cloud leader allows to leaving node to leave the network. After that cloud leader update the table for new upcoming member.

### E. RELEASE THE CLOUD:

When cloud leader moves out of range of network or no need of cloud resources then it will release the resources so that other vehicles can also use the vehicular cloud. For this purpose, the cloud leader sends the message to all vehicles that are participant of network and release all resources of clouds.

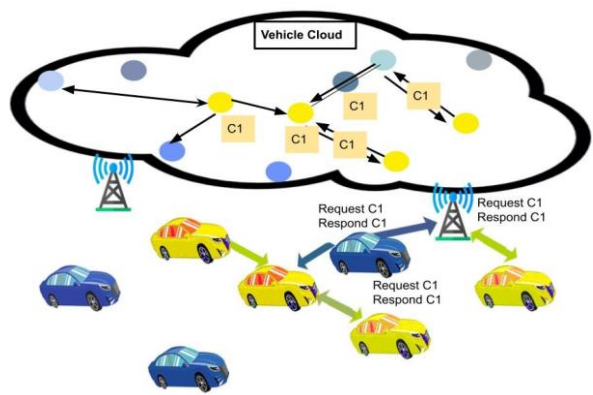

**Fig 9:** Architecture of IOV based on vehicular clouds

A Systematic Literature Review is focused on the research that has been conducted on the use of IoV technology to address traffic congestion on roads. IoV architecture and communication protocols and taxonomy of Traffic congestion detection and prediction techniques shown in Figure 9, Congestion avoidance and traffic routing strategies, IoV-based traffic management and control systems and Evaluation and performance analysis of IoV-based traffic management systems.

**Comparison:**

Table 8: Comparatively analysis between VANET and IOV

| Sr# | TRADITIONAL VANET | IOV based on Cloud computing |
|---|---|---|
| 1 | Infrastructure-less | Infrastructure based |
| 2 | No real time Information sharing | Real time information |
| 3 | Covered less area | More covered Area |

| 4 | MANET based | IOT Based |
|---|---|---|
| 5 | More security threats | Less security threats than VANET |
| 6 | Self-resources | Complex Algorithm |
| 7 | Self-resources | Clouds based resources |

By analysis the table 8 we have found many key differences between VANET and IOV and we can explain all reason to best IOV architecture due to main points mentioned in the table 8. VANET (Vehicular Ad Hoc Network) and IOV (Internet of Vehicles) are two distinct concepts, but they are closely related to the advancement of technology in transportation.

VANET is a wireless network that allows communication between vehicles and between vehicles and road infrastructure such as traffic lights, sensors, and cameras. The primary objective of VANET is to provide safety and traffic management on the road by exchanging information about accidents, traffic congestions, and weather conditions, among others. IOV, on the other hand, is a network of vehicles and other objects such as traffic lights, cameras, and sensors that are interconnected through the Internet. Comparative analysis among previous study and modern vehicle system of IoV in figure 9. The primary goal of IOV is to create an ecosystem of vehicles that can share information and provide a range of services, such as infotainment, vehicle diagnostics, and remote assistance.

## VI. FUTURE WORK

### A. CONTRIBUTION

We have selected and read many articles in term of get the answers to set questions for SLR. Our main goal to point out modern vehicular control system beneficial strategies. IOV can be implemented in many ways but we have select the Vehicular cloud computing. The reason behind this strategy is that no extra central control system is required. Just pick leader capability vehicle which can be create vehicular model then task assign to different capable vehicles to collect traffic related information after that leader vehicle execute and share all information to whole traffic of city. In this network the main role is IoT in which all stakeholders devices is connected in term of create intelligent traffic control system. We have set many Questions to collect this information from many articles. Every answers of every question is to related comparatively better performance than traditional one. At the end we are also point out some challenges which may happened in IOV architectures. And we are need to deal to this challenges for better performance. The IOV (Internet of Vehicles) traffic control system has numerous benefits for drivers, pedestrians, passengers, and the city's overall infrastructure. However, it also poses some challenges that need to be addressed. One of the major issues is security threats, as well as communication failures or delayed communication due to the dynamic running protocols of moving vehicles. To overcome these challenges, further research should be conducted to improve the network. One potential solution could be to implement 5G network technology, which would enhance the system's capabilities. In addition, installing modern devices that can guide drivers to park their vehicles correctly and prevent them from wrong parking and over-speeding in sensitive areas could be helpful. By addressing these challenges, the IOV traffic control system can continue to provide significant benefits to the community.

### B. CHALLENGES

This system may face many problems that should be resolve in term of best IOV architecture.

- There is need to create and maintain all those algorithms and protocols which should be better opera table the vehicle network. Vehicles are constantly moving and change the distance between them. So, protocol must be able of dynamic fashion of networking. [9]

- This system may face problems in term of intelligent routing and path planning.

- Real time huge amount of data may process on the specific node of network. If this processing slows down the real time communication system may affect.

- All participants of Cloud computing need security form vulnerable threats. All components like central communication system vehicles, wrong path information, error SUMO running software on current time or may slow down.

- Denial of services (Do's) may happened on any vehicle then this node can be share traffic information to others or leader vehicle. This may slow down the network.

- Denial of services may Couse the jamming problem in medium in which is used for commonly sharing information. And there may also threats like message declaration, message refuse. For security system we need proper authentication, integrity control system.

- There may be privacy information issues may happen if hackers are involved between vehicular to road side unit (RSU) to vehicular to vehicular (V2V) or vehicular to pedestrians. This is need proper authentication need for which information should be share and which privacy information of vehicle remain.

- There may be other challenge of service delay can be faced in some situation where traffic is too huge amount. This challenge is mostly due to continuously changing the location of vehicles, dynamic routing protocol may face problem in some situations. Proper location sharing is mostly through GPSs system. Current location sharing is very important for avoidance the collisions. So proper dynamic protocol system is very important in this situation.

- There is important to control IOV information security system to deal with attackers, control communication security system holes like diversity traffic, privacy contravention, changing the messages, stop or keep away from using the cryptographic algorithm based software.

## VI. CONCLUSION

The systematic literature review indicates that extensive research has been conducted on the topic of the internet of vehicles (IoV) from 2014 to 2023. IoV has emerged as a potential solution to tackle traffic congestion on roads, as evidenced by multiple studies demonstrating its ability to enhance traffic flow, reduce congestion, and improve safety. By collecting data from vehicles, infrastructure, and sensors, IoV systems can provide real-time traffic information to drivers and traffic management authorities, which can be utilized to optimize traffic flow and alleviate congestion. The literature review highlights several approaches and techniques, including machine learning, data analytics, and artificial intelligence, that can be leveraged to develop IoV-based traffic congestion systems. These techniques can help to improve the accuracy of traffic predictions, which can be used to make better traffic management decisions. Overall, the literature review highlights the potential of IoV-based traffic congestion systems to address the problem of traffic congestion on roads. However, it is important to note that there are still several challenges that need to be addressed, such as data privacy, cybersecurity, and interoperability. Therefore, future research should focus on addressing these challenges to ensure the effective implementation of IoV-based traffic congestion systems. Our work was about to conduct SLR in term of best solution to deal with road congestion issues in big cities. We have chosen the technique to explore the problems in traditional VANET system. This outcomes of SLR are very effective to explore issues and we have also elaborated the technique of new architecture by which most of issues were solved. System is based on IOV based on vehicular cloud computing to control traffic related problems. We have found by this SLR that this system has solve many traffic issues and we need to apply this technology in all big cities. But there were few challenges in new architecture which we have found by literature. We need to deal better of these challenges. But overall, this system is very effective to control the traffic related issues.